\documentclass[5p,compress,numbers]{elsarticle}
\usepackage{epstopdf}
\usepackage{subfigure,graphicx,tabularx}
\usepackage{float}
\usepackage{amsmath, amsfonts, amssymb,mathrsfs}
\usepackage{amsthm}
\usepackage{epsf}
\usepackage{amsfonts}
\usepackage{stmaryrd}
\usepackage{amssymb}
\usepackage{leftidx}
\usepackage{color}
\usepackage{mathtools}
\usepackage{placeins}
\usepackage{booktabs}
\usepackage{enumitem}
\usepackage{caption}
\usepackage{multirow}
\usepackage{stackengine}
\usepackage[utf8]{inputenc}
\usepackage[english]{babel}
\usepackage{color}
\usepackage{pifont}
\usepackage{fontawesome}
\newcolumntype{?}{!{\vrule width 1pt}}

\theoremstyle{plain}
\newtheorem{theorem}{Theorem}

\newtheorem{remark}[theorem]{Remark}

\def\bfu{{\bf u}}

\def\bfE{{\bf E}}

\def\bfS{{\bf S}}

\def\bfX{{\bf X}}

\def\bfe{{\bf e}}

\newcommand\sts{s_{\texttt{ts}}}

\begin{document}
\begin{frontmatter}

\title{Fracture under body forces: An accessible test and analysis}

\author{Farhad Kamarei}
\ead{kamarei2@illinois.edu}

\author{Sarah Bueno De Castro}
\ead{sarahbc2@illinois.edu}

\author{Jeffery R. Roesler}
\ead{jroesler@illinois.edu}

\author{Oscar Lopez-Pamies}
\ead{pamies@illinois.edu}

\address{Department of Civil and Environmental Engineering, University of Illinois, Urbana--Champaign, IL 61801, USA }

\begin{abstract}

\vspace{-0.1cm}

Fracture in large-scale structures --- from failing concrete dams to calving glaciers --- is dominated by self-weight. Yet, owing to their immense scale, directly studying when and where cracks nucleate and propagate in these systems under gravity is unfeasible. To date, experimental investigations have largely relied on mimicking self-weight in scaled-down specimens via centrifugal forces, necessitating specialized centrifuge facilities. In this Letter, we introduce a simple benchtop test that bypasses the need for this infrastructure. The method utilizes a cantilever beam containing a downward-pointing V-notch, loaded solely by its self-weight. Through appropriate selection of the notch geometry and placement and the beam dimensions, the setup forces a crack to nucleate at the notch tip and propagate through the specimen in a measurable manner. We demonstrate the utility of this self-weight fracture test on a 3D-printable mortar and analyze the results using the fracture theory recently introduced in (\citeauthor{LPKFG2026}, 2026). \emph{Inter alia}, the findings provide a first validation of the theory for fracture driven by body forces.

\vspace{0cm}

\keyword{Strength, Toughness, Flow and Fracture, Energy Methods}
\endkeyword

\end{abstract}

\end{frontmatter}

\section{Introduction}\label{Sec: Intro}

The recognition that gravity plays a dominant role in the fracture of large-scale structures --- whether man-made, such as dams, or natural, such as glaciers --- dates back to the early work of Galileo \cite{Galileo} in the 1630s. Despite this early realization, the massive size of these structures has long rendered direct experimental fracture studies unfeasible. A workaround emerged in the early 1990s by leveraging the correspondence between gravitational and centrifugal forces \cite{Broz1991,Plizzari1995}. Specifically, this approach simulates the effects of gravity by applying centrifugal forces to scaled-down replicas of the structures of interest --- a technique pioneered decades earlier to study large rock structures \cite{Bucky1935}. Alongside this, other techniques have sought to approximate gravity on scaled models by applying a distribution of discrete boundary loads \cite{Carpinteri1992}.

In this Letter, we introduce a new test aimed at directly studying fracture driven by gravitational body forces. As elaborated in Section \ref{Sec: Test}, the test consists of a beam of square cross section that contains a downward-pointing V-notch, clamped as a cantilever. By appropriately choosing the geometry and placement of the notch and the dimensions of the beam, a crack is forced to nucleate from the notch tip and subsequently propagate, severing the beam. In contrast to the techniques mentioned above, the proposed test relies strictly on actual gravity to drive fracture and entirely bypasses the need for specialized testing infrastructure. In Section \ref{Sec: Results}, the utility of this self-weight fracture test is demonstrated on a 3D-printable mortar. The results are then analyzed in Section \ref{Sec: Theory} by means of the theory of fracture nucleation and propagation recently introduced in \cite{LPKFG2026}. Finally, Section \ref{Sec: Final Comments} provides several concluding remarks.

\section{The test}\label{Sec: Test}

\begin{figure}[t!]
\centering
\includegraphics[width=0.9\linewidth]{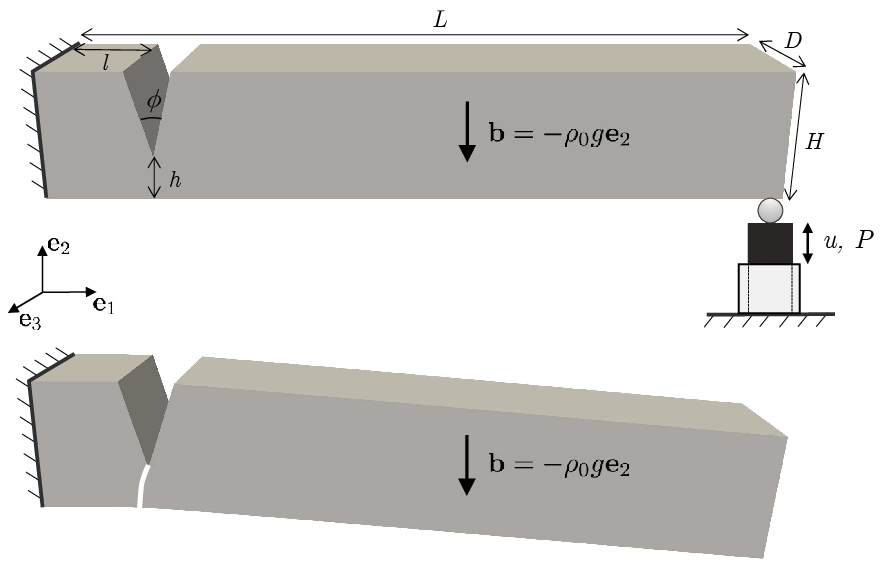}
\caption{{\small Schematic of the proposed test: (top) initial undeformed configuration showing the applied gravitational body force and the retractable lower support; (bottom) fractured state following crack nucleation at the notch tip and subsequent propagation.}}\label{Fig1}
\end{figure}

The rationale for the proposed test is based on two guiding principles. First, the procedure must be practical for both laboratory and field environments, requiring only standard equipment (e.g., displacement controlled hydraulic actuators and load cells). Second, the fracture process must be driven by actual gravity. With these in mind, we initially evaluated both beams and plates utilizing various combinations of cantilever and simple supports to create regions where self-weight naturally induces localized tension. Ultimately, due to its overall practicality in terms of specimen fabrication, positioning into the holding frame, and testing, we chose the setup schematically depicted in Fig.~\ref{Fig1}.

Generically, the specimen consists of a beam of length $L$, height $H$, and depth $D=H$, containing a V-notch with angle $\phi$ located at a distance $l$ from the left boundary and a distance $h$ from the bottom edge. The beam is fully clamped at its left boundary to form a cantilever and is subjected to the body force of gravity, which in our frame of reference (per unit undeformed volume) reads $\textbf{b}=-\rho_0 g \bfe_2$, where $\rho_0$ is the material mass density and $g$ is the gravitational acceleration. Because gravitational acceleration is constant ($g=9.81$~m/s$^2$), the right free end of the beam is initially supported from below. This support is gradually lowered, smoothly transitioning the beam to a state where it is loaded solely by its self-weight. By properly selecting the geometry and placement of the notch and the dimensions of the beam, one can identify a setup such that a crack nucleates from the notch tip and propagates toward the bottom edge, severing the beam, in a manner that can be monitored and measured experimentally.

%
%

\subsection{Choosing appropriate notch and beam geometries}

\paragraph{The notch} To force crack nucleation and propagation to take place in a confined, easily monitored region around the notch tip, the distance $h$ from the notch tip to the bottom edge should be chosen as the smallest macroscopic length scale in the system. To prevent the specimen from becoming unnecessarily heavy, and hence more difficult to handle, the value of $h$ should be minimized while remaining ``much larger'' than the maximum microstructural heterogeneity of the material, say $\ell$. This guarantees an adequate separation of length scales, ensuring that the local region where the crack nucleates and propagates can be treated as a homogeneous macroscopic subdomain. 

Currently, the threshold for what constitutes ``much larger'' in terms of fracture behavior has not been fully settled. For isotropic linear elastic properties --- the Young's modulus $E$ and Poisson's ratio $\nu$ --- it suffices to set $h$ to be four times larger than the maximum microstructural heterogeneity \cite{DW1996}. For isotropic strength (the strength surface\footnote{Throughout this work, $\bfS$ denotes the first Piola-Kirchhoff stress tensor and $s_1\geq s_2 \geq s_3$ are the principal nominal stresses.} $\mathcal{F}(\bfS)=\widehat{\mathcal{F}}(s_1,s_2,s_3)=0$) and fracture toughness ($G_c$), a limited number of experiments indicate that $h$ should be at least ten times larger, possibly more \cite{Gonnerman1925,Blanks1935,Wittmann1988,Wittmann1998}. By way of example, testing a polycrystalline ceramic with a $10~\mu$m grain size would require $h \geq 100~\mu$m; testing polycrystalline ice with $1$~mm grains would require $h \geq 10$~mm; testing mortar with a maximum aggregate size of $4.75$~mm would require $h \geq 47.5$~mm; and testing concrete with a maximum aggregate size of $25$~mm would require $h \geq 250$~mm.

To further limit the overall weight of the specimen, while guarding against nucleation at the clamped boundary and ensuring the crack expected to nucleate from the notch tip does not propagate into the clamp, the distance $l$ from the notch to the clamped boundary should be chosen as the second smallest macroscopic length scale in the system.

Finally, the notch angle $\phi$ should be chosen to be relatively sharp to promote early nucleation, without compromising manufacturability or being so acute that it effectively acts as a pre-existing crack. The objective is to study the richer phenomena of crack nucleation from a region of high stress concentration, followed by propagation, rather than simply the growth of a pre-existing crack.

\paragraph{The beam} The height $H$ (and depth $D=H$) of the beam must be chosen sufficiently larger than $l$. In turn, because the bending stress near the clamped support, where the notch is located, scales with $L^2/H$, the length $L$ should be significantly larger than the height $H$ to ensure that self-weight generates sufficient stress for crack nucleation. Concurrently, the overall volume $V=L H^2 - H(H-h)^2 \tan(\phi/2)$ must be minimized to maintain a manageable specimen mass.

\medskip

Collectively, the above competing constraints establish the following hierarchy of material and specimen length scales:
\begin{equation}\label{hierarchy}
\ell \ll h < l < H = D < L.
\end{equation}
The selection of precise values for $h$, $l$, $H$, and $L$ that satisfy these constraints depends on the properties of the material being tested and the local gravitational acceleration $g$. Ultimately, these values must be selected to ensure that fracture does nucleate at the notch tip while keeping the weight of the specimen as small as possible. 

Following \cite{LPKFG2026}, to ensure that fracture does nucleate at the notch tip, the strength surface of the material, $\mathcal{F}(\bfS)=0$, must be exceeded ($\mathcal{F}(\bfS)\geq0$) over a sufficiently large region around the notch when the support is fully retracted and the beam is subject to its entire self-weight. To identify this region of strength violation, one must compute the stress field in the specimen and evaluate where $\mathcal{F}(\bfS)\geq0$. This is readily accomplished via standard finite-element (FE) analysis.\footnote{All FE simulations presented in this and the next two sections were carried out with quadratic tetrahedral elements.} As a preliminary indicator of fracture, the height of this region must be in the order of $3EG_c/(8\sts^2)$, where $\sts$ stands for the uniaxial tensile strength of the material \cite{SRLP26}. However, definitive confirmation requires the type of analysis detailed in Section \ref{Sec: Theory}. In the following subsections, we demonstrate this procedure for two materials --- a 3D-printable mortar, pertinent to large-scale man-made structures, and ice, pertinent to natural ones --- thereby determining the specific geometric parameters $h$, $l$, $\phi$, $H$, and $L$.

\subsection{An example: A 3D-printable mortar}

We begin by considering the class of 3D-printable mortars recently introduced in \cite{Manaugh2025} and studied in \cite{SMMRLP26}. In the present study, the specific mixture consists of Type IL cement with a water-to-cement weight ratio of 0.38, 50\% volume fraction of natural sand, and no admixtures. Based on the suite of experiments introduced in \cite{SMMRLP26}, this mixture can be described, macroscopically, as a homogeneous isotropic linear elastic brittle material with constant mass density $\rho_0$, isotropic elastic energy density 
\begin{equation}\label{W-Lin-elastic}
W(\bfE)=\dfrac{E}{2(1+\nu)}\operatorname{tr}(\bfE^2)+\dfrac{E\nu}{2(1+\nu)(1-2\nu)}(\operatorname{tr}\bfE)^2, 
\end{equation}
Drucker-Prager strength surface 
\begin{equation}
\mathcal{F}(\bfS)=\sqrt{\mathcal{J}_2}+\dfrac{s_{\texttt{cs}}-s_{\texttt{ts}}}
{\sqrt{3}\left(s_{\texttt{cs}}+s_{\texttt{ts}}\right)}\, \mathcal{I}_1-\dfrac{2s_{\texttt{cs}} s_{\texttt{ts}}}
{\sqrt{3}\left(s_{\texttt{cs}}+s_{\texttt{ts}}\right)}=0,\label{DP}
\end{equation}
scalar fracture toughness $G_c$, and the material constants after seven days of moist curing listed in Table~\ref{Table1}. In these expressions, $\bfE(\bfu)=\frac{1}{2}(\nabla\bfu+\nabla\bfu^T)$ denotes the symmetrized gradient of the displacement field $\bfu$, or infinitesimal strain tensor, $\mathcal{I}_1=s_1+s_2+s_3$, $\mathcal{J}_2=\frac{1}{3}(s^2_1+s^2_2+s^2_3-s_1 s_2-s_1 s_3-s_2 s_3)$, while, again, $s_{\texttt{ts}}$ and $s_{\texttt{cs}}$ stand for the uniaxial tensile and compressive strengths. That is, $s_{\texttt{ts}}$ and $s_{\texttt{cs}}$ denote the critical nominal stress values at which fracture nucleates under uniform states of monotonically increased uniaxial tension $\bfS={\rm diag}(s>0,0,0)$ and compression $\bfS={\rm diag}(-s<0,0,0)$, respectively.
\begin{table}[b!]\centering\small
\caption{Material constants (after seven days of moist curing) for the 3D-printable mortar studied in this work.}
\begin{tabular}{l? c c }
\toprule
Mass      & $\rho_0$ (kg/m$^3$)     &        \\
density   & $2210$                  &           \\
\midrule
Elasticity&   $E$ (GPa) & $\nu$ \\
constants &   $26.6$      & 0.17  \\
\midrule
Strength  & $s_{\texttt{ts}}$ (MPa) & $s_{\texttt{cs}}$ (MPa)       \\
constants & $3.5$                     & $45$                      \\
\midrule
Fracture  & $G_c$ (N/m)             &       \\
toughness & $20$                    &         \\
\bottomrule
\end{tabular} \label{Table1}
\end{table}

Given that the largest underlying aggregate in the mortar is $\ell=4.75$ mm in diameter, we set $h=0.1$ m. This corresponds to a separation-of-length-scales factor of $h/\ell=21.1$. In turn, in view of the inequalities (\ref{hierarchy}), we choose $l=0.2$ m, $H=D=0.3$ m, and $\phi=30^\circ$. It remains to select a specific value for $L$ consistent with (\ref{hierarchy}) to ensure that fracture nucleates at the notch tip while minimizing the weight of the specimen. 

To this end, we note that the formula $S_{EB}=3\rho_0 g (L^2-l^2)H/h^2$ from the Euler-Bernoulli beam approximation provides an estimate for the maximum principal stress $s_1$ around the notch tip. Equating this formula to the uniaxial tensile strength $\sts$ yields a length of $$L=\sqrt{l^2+\sts h^2 / (3 \rho_0 g H)}=1.35\, \text{ m}.$$ Having determined this estimate, the next step is to perform an elastic FE analysis of the regions where the strength surface is violated in a beam of the above dimensions, with a length somewhat larger than 1.35 m, under self-weight. We choose $L=1.7$ m. 

\begin{figure}[t!]
\centering
\includegraphics[width=0.99\linewidth]{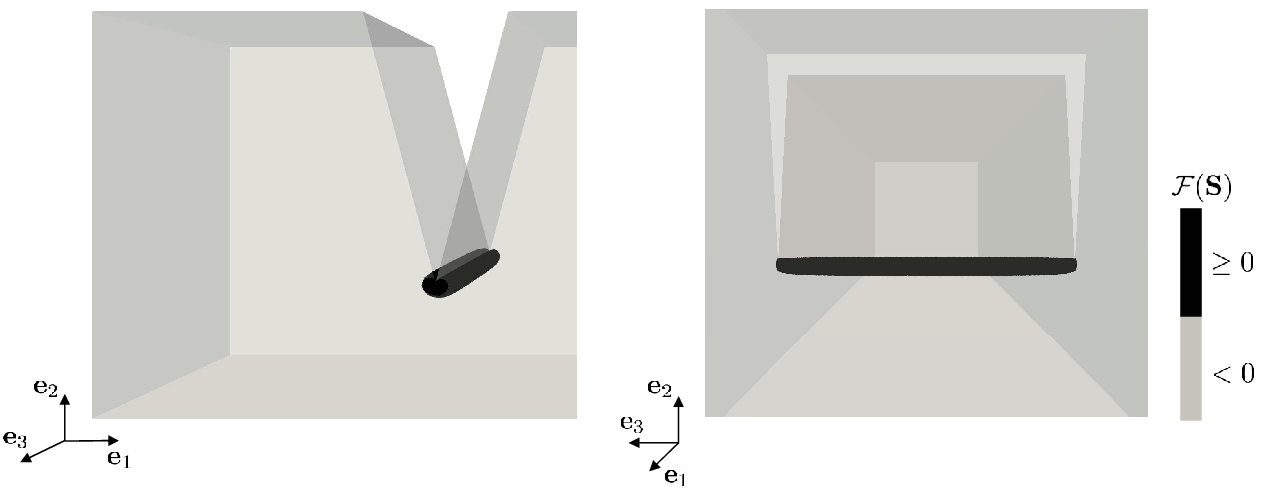}
\caption{{\small Contour plots from two perspectives showing the region around the notch where the stress field exceeds the strength surface of the material ($\mathcal{F}(\mathbf{S})\geq0$). These results correspond to a mortar beam, with dimensions $L=1.7$ m, $H=D=0.3$ m, $l=0.2$ m, $h=0.1$ m, $\phi=30^\circ$, properties listed in Table \ref{Table1}, subjected to self-weight ($g=9.81$ m/s$^2$).}}\label{Fig2}
\end{figure}

Based on a standard FE analysis, Fig. \ref{Fig2} shows two magnified perspectives of the region around the notch tip where the strength surface is indeed violated, $\mathcal{F}(\mathbf{S}) \geq 0$, as expected. We emphasize that the strength surface is also violated around the edges of the beam at the clamped end, but those regions are highly localized and not visible in Fig.~\ref{Fig2}. The height of the region of strength violation is almost constant through the depth, increasing from $12.3$ mm at the back and front boundaries to $12.7$ mm at the center of the beam. Following the analysis for beams presented in \cite{SRLP26}, we note that this height is in the order of the material length scale $3EG_c/(8\sts^2)=16.3$ mm. Hence, fracture nucleation is expected. The analysis presented in Section \ref{Sec: Theory} confirms that the region of strength violation is indeed large enough for fracture to occur, thereby validating our choice of $L=1.7$ m for the beam length. Note that this length leads to a specimen of mass $m=\rho_0 H(L H - (H-h)^2 \tan(\phi/2))=331$ kg, which is significant but certainly manageable in both laboratory and field environments.

%
%

\subsection{Another example: Ice}

As a complementary example, this subsection repeats the preceding analysis to establish appropriate values for $h$, $l$, $\phi$, $H$, and $L$ for ice.

Due to its ubiquity in nature, ice has been studied extensively \cite{SchulsonDuval2009}. Its microstructure and macroscopic mechanical behavior are highly dependent on how it is formed and the conditions under which it is tested. Here, we consider the typical case of isotropic polycrystalline freshwater ice with an average grain size of $1$~mm. Assuming it is tested fast enough that it can be considered a homogeneous, isotropic, linear elastic brittle material, we characterize it using a constant mass density $\rho_0$, the isotropic elastic energy density (\ref{W-Lin-elastic}), the Drucker-Prager strength surface (\ref{DP}), a scalar fracture toughness $G_c$, and the material constants listed in Table~\ref{Table2}.

With the grains measuring $\ell=1$ mm in average diameter, we set $h=0.05$ m to be fifty times this microstructural length. In accordance with (\ref{hierarchy}), the other dimensions are set to $l=0.1$ m, $H=D=0.2$ m, and $\phi=30^\circ$. In turn, from the Euler-Bernoulli beam estimate, we get $L=\sqrt{l^2+\sts h^2/(3 \rho_0 g H)}=0.84$ m. In view of this result, we set $L=1$ m for the beam length. This results in a specimen mass of $m=\rho_0 H(L H -(H-h)^2 \tan(\phi/2))=36$ kg, which is manageable, even more so than that of the 3D-printable mortar.

A standard FE analysis shows that the height of the region of strength violation is nearly constant across the depth, increasing only from $5.9$ mm at the boundaries to $6$ mm at the center. This height is greater than the material length scale $3EG_c/(8\sts^2)=3$ mm. Therefore, fracture nucleation is expected. An analysis mirroring the one presented in Section \ref{Sec: Theory} verifies that this strength-violation region is indeed large enough for fracture to occur, thereby validating $L=1$ m as an appropriate beam length.

\begin{table}[t!]\centering\small
\caption{Material constants for ice.}
\begin{tabular}{l? c c }
\toprule
Mass      & $\rho_0$ (kg/m$^3$)     &        \\
density   & $920$                  &           \\
\midrule
Elasticity&   $E$ (GPa) & $\nu$ \\
constants &   $9$      & 0.325  \\
\midrule
Strength  & $s_{\texttt{ts}}$ (MPa) & $s_{\texttt{cs}}$ (MPa)       \\
constants & $1.5$                     & $10$                      \\
\midrule
Fracture  & $G_c$ (N/m)             &       \\
toughness & $2$                    &         \\
\bottomrule
\end{tabular} \label{Table2}
\end{table}

\section{Experimental results for the 3D-printable mortar}\label{Sec: Results}

We now present experimental results for the proposed test, conducted on the 3D-printable mortar using the specimen dimensions established above: $h=0.1$ m, $l=0.2$ m, $\phi=30^\circ$, $D=H=0.3$ m, and $L=1.7$ m.

%
\begin{figure}[b!]
\centering
\includegraphics[width=0.95\linewidth]{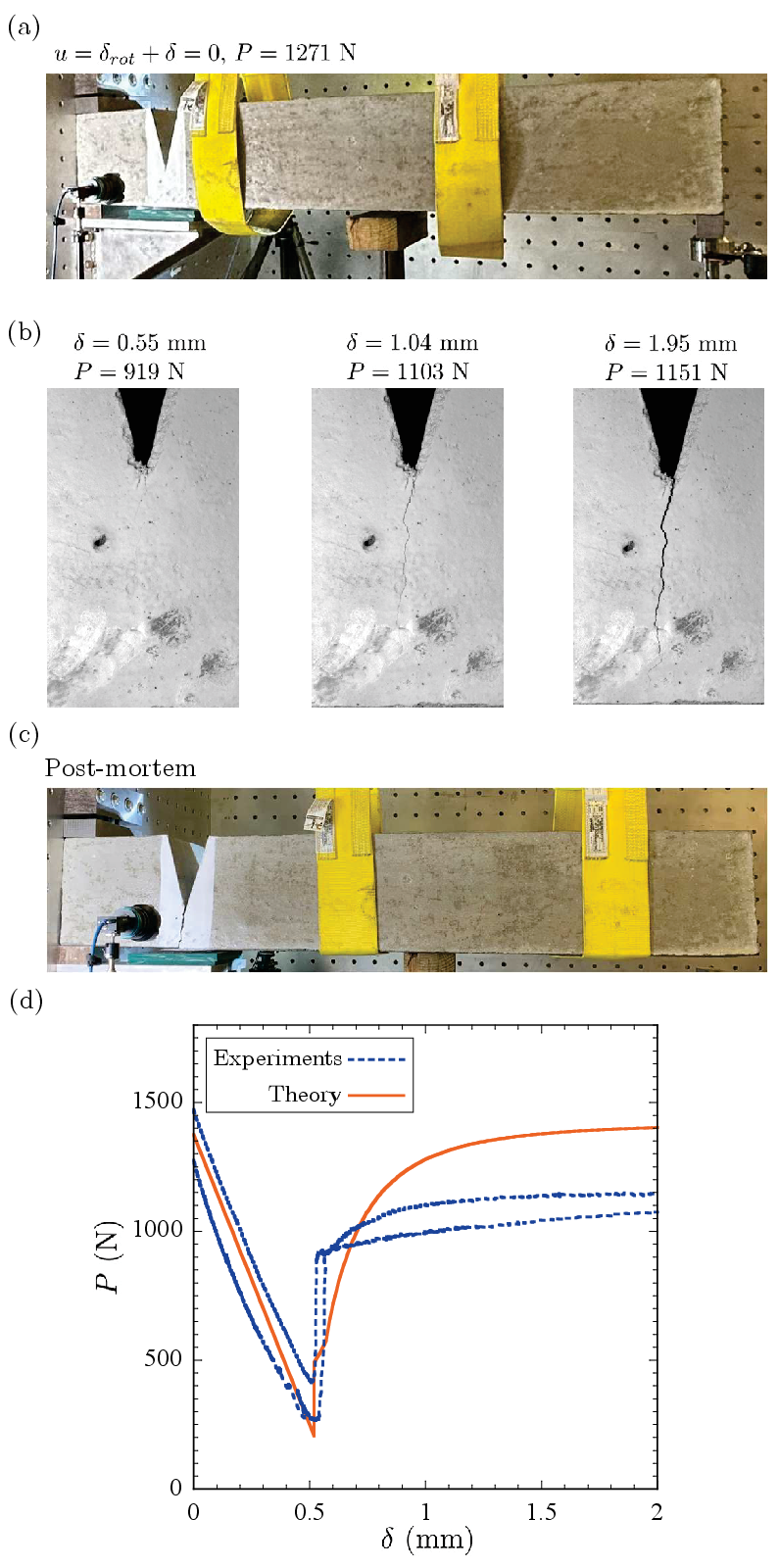}
\caption{{\small (a) Initial configuration and (b) front views of the notch-tip region at three intermediate stages of support retraction, annotated with the corresponding deflection $\delta$ and support force $P$ for one of the specimens. (c) Post-mortem configuration of the same specimen. (d) Measured (dashed lines) and predicted (solid line) $P$ vs.~$\delta$ responses for the two tests.}}\label{Fig3}
\end{figure}
%

We carried out two tests on two different specimens. Each specimen was cast in a mold featuring an extra $0.1$ m of length to the left of the notch. This added section was designed to slide between two steel blocks, which were then tightened to fix the beam as a cantilever; see Fig.~\ref{Fig3}(a). 
\begin{remark}\label{Remark: Non-Prismatic}
\emph{The geometry of both specimens was measured after casting and demolding. Rather than being perfectly prismatic, both exhibited slightly different varying depths ($0.301 \text{ m} \leq D \leq 0.309 \text{ m}$) near the notch. These geometric variations have a measurable impact on the gravitational body forces experienced by the specimens, and consequently, on the final results.}
\end{remark}
The support at the free end consisted of a steel bar, of width $w=2$ cm, spanning the entire depth of the specimen. This plate rested on an actuator capable of being lowered with a displacement resolution of $25~\mu$m, which was also equipped with a load cell featuring a resolution of $5$ N; see Fig.~\ref{Fig3}(a). 

To monitor the expected fracture nucleation at the notch tip and its subsequent propagation, two cameras were utilized to optically image the front and back of the specimen, focusing on the notch region.

After the specimen was clamped between the steel blocks to enforce a fixed cantilever condition and verified to be level, the test was initiated by retracting the support at a rate of $r=0.25 \,\mathrm{mm/min}$ while recording three outputs as functions of support displacement $u$: (\emph{i}) front and (\emph{ii}) back optical images of the notch region, and (\emph{iii}) the reaction force $P$ at the support.
\begin{remark}
\emph{Because compliance at the clamped support induces a rigid-body rotation, the applied support displacement $u$ comprises both an apparent deflection from this rotation, $\delta_{\mathrm{rot}}$, and the true flexural deflection of the beam, $\delta$. Given the specimen length $L = 1.7\text{ m}$, even a slight rotation of $\theta_{\mathrm{rot}} = 0.05^\circ$ produces a substantial apparent deflection $\delta_{\mathrm{rot}} = L \tan\theta_{\mathrm{rot}} = 1.48\text{ mm}$. Prior to the nucleation of fracture, however, the response is governed by linear elastostatics. Therefore, $\delta_{\mathrm{rot}}$ can be identified by comparing the experimental load-displacement curves against the corresponding FE simulations up to the displacement at fracture nucleation, $u_n$. Across the two tests conducted, this procedure yields
$$\delta_{\mathrm{rot}} = \begin{cases}  0.749 u + 0.022 u^2, & 0 \leq u \leq u_n \\ 0.749 u_n + 0.022 u_n^2, & u > u_n \end{cases},$$
reflecting the observation that the rigid-body rotation remains essentially frozen after fracture nucleates.}
\end{remark}

Figure \ref{Fig3} presents images of one of the specimens at various stages of testing: its initial configuration ($u =\delta_{rot}+\delta=0$, $P = 1249$ N), the notch-tip region at three intermediate levels of support displacement $u$ (with corresponding values for the deflection $\delta$ and force $P$), and its final post-mortem state. Additionally, the figure plots the measured support force-deflection responses for both specimens, alongside the theoretical prediction detailed in the following section.

The results yield three main observations. First, within the aforementioned specimen-to-specimen variability associated with fabrication and setup, the tests demonstrate reproducibility. Second, fracture nucleation occurs at the notch tip when the support force $P$ reaches a minimum in the $P$ vs.~$\delta$ response. For the two specimens, these minimum values are $P = 276$~N and $410$~N, which are attained at $\delta = 0.48$~mm and $0.51$~mm, respectively. As the support is lowered further, the crack propagates stably along its initial nucleation angle of approximately $-5^\circ$ relative to gravity, eventually severing the beam; during this stable propagation phase, the support force $P$ increases. Third, as detailed in the next section, the theoretical predictions are in good agreement with the experimental observations of when and where the crack nucleates and propagates.

\section{Theoretical analysis of the tests on the 3D- printable mortar}\label{Sec: Theory}

In a recent contribution, Lopez-Pamies et al. \cite{LPKFG2026} have introduced a macroscopic, or continuum, theory that describes when, where, and why cracks nucleate and propagate in nominally elastic brittle materials under monotonic, quasi-static, but otherwise arbitrary mechanical loads. The theory posits that \emph{cracks nucleate and propagate exclusively in regions where the strength surface of the material is exceeded, with their evolution dictated by the minimization of the sum of the potential energy (the elastic energy minus the work done by externally applied forces) and the surface energy}. Comparisons with the ``Nine Circles of Elastic Brittle Fracture'' \cite{KZDLP26} --- nine tests that span the entire range of well-settled experimental knowledge on fracture nucleation and propagation in hard and soft materials --- suggest that the theory accurately describes reality. In this section, we work out its predictions and compare them with the above experiments on the 3D-printable mortar.

For the problem of interest here, denoting by $\Omega$ the domain occupied by the specimen in its undeformed and stress-free configuration, by $\Gamma$ the unknown crack set, and by $\overline{\bfu}(\bfX,t)$ the time-dependent displacement prescribed on the Dirichlet part $\partial\Omega_\mathcal{D}=\partial\Omega_{\texttt{clamp}}\cup\partial\Omega_{\texttt{support}}$ of the boundary $\partial\Omega$, with $\partial\Omega_{\texttt{clamp}}=\{\bfX: X_1=0,\,|X_2|<H/2,\,|X_3|<D/2\}$ and $\partial\Omega_{\texttt{support}}=\{\bfX: L-w/2<X_1<L,\,X_2=-H/2,\,|X_3|<D/2\}$, we write
\begin{multline*}
\mathcal{P}(\bfu,\Gamma):=\int_{\Omega\setminus\Gamma}W(\bfE(\bfu(\bfX)))\,{\rm d}\bfX+
\rho_0 g\displaystyle\int_{\Omega\setminus\Gamma}u_2(\bfX)\,{\rm d}\bfX
\end{multline*}
for the potential energy. The theory states that, at any discrete time $t_k\in\{0=t_0 <t_1,...<t_m<t_{m+1},...<t_M=T\}$, the pair displacement-field-crack-set $(\bfu_k=\bfu_k(\bfX),$ $\Gamma_k)$ minimizes
\begin{align}
\mathcal{E}(\bfu,\Gamma):=\mathcal{P}(\bfu,\Gamma)+G_c\,\mathcal{H}^{2}\left(\Gamma\right)\label{Fracture-Energy-Min}
\end{align}
among all $(\bfu,\Gamma)$ with

\begin{equation*}
\left\{\begin{array}{l}
\bfu(\bfX)=\overline{\bfu}(\bfX,t_k)=\left\{\hspace{-0.1cm}\begin{array}{ll} \textbf{0}, & \bfX\in \partial\Omega_{\texttt{clamp}}\\[2mm]
-u(t_k)\,\bfe_2, & \bfX\in \partial\Omega_{\texttt{support}}\end{array}\right. \\[4mm] 
\bfu(\bfX) \textrm{ satisfies material impenetrability} \\[2mm] 
\Gamma\setminus\Gamma_{k-1}\subset \mathcal{V}_{\mathcal{F}}(t_{k-1})\\[2mm] 
\Gamma\supset\Gamma_{k-1} \end{array}\right. .
\end{equation*}
In these expressions, $\mathcal{H}^{2}(\Gamma)$ denotes the $2$-dimensional Hausdorff measure  (the surface measure) of the unknown crack $\Gamma$ in the undeformed configuration, $u(t)=r t$ with $r$ denoting the rate at which the support is retracted, and
\begin{equation*}
\mathcal{V}_{\mathcal{F}}(t)=\{\bfX\in\Omega\setminus\Gamma: \mathcal{F}(\bfS(\bfX,t))\ge0\}
\end{equation*}
is the set of all material points at which the strength surface $\mathcal{F}(\bfS)=0$ of the material is exceeded at time $t$, referred to as the strength violation set. 

%
\begin{figure}[b!]
\centering
\includegraphics[width=0.95\linewidth]{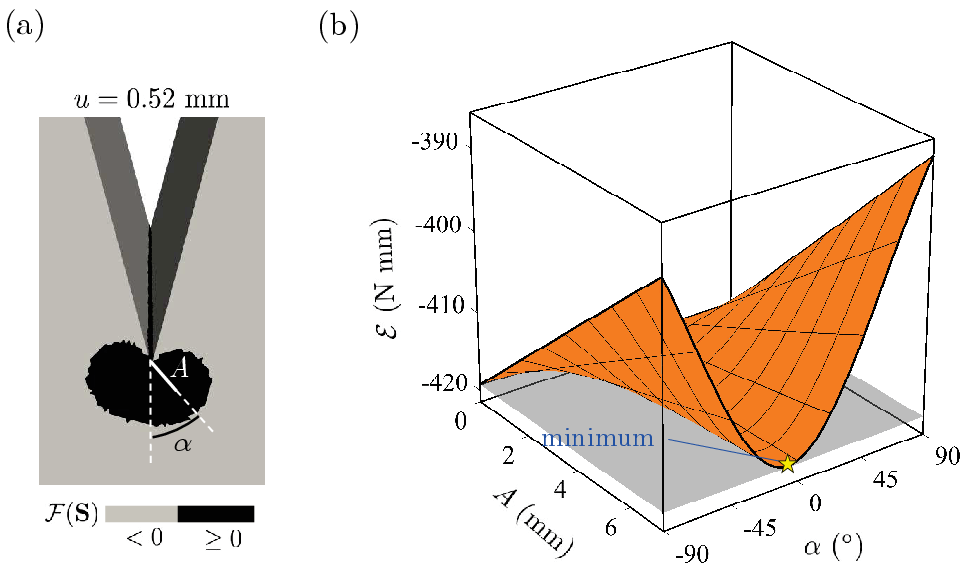}
\caption{{\small Elastic simulations of the test, at the support displacement $u=0.52$ mm, on mortar containing test cracks of length $A$, thickness $D=0.3$ m, oriented at angles $\alpha$ with respect to the direction of the gravitational body force. (a) Contour plot of a magnified region of the specimen where the stress field exceeds ($\mathcal{F}(\bfS)\geq0$) the strength surface of the material, depicting the test cracks considered within. (b) The energy $\mathcal{E}$ of the specimen plotted as a function of the test crack length $A$ and angle $\alpha$.}}\label{Fig4}
\end{figure}
%

The pertinent calculations are as follows. We begin by numerically (again, via FE) identifying the regions where the strength surface is violated, $\mathcal{F}(\mathbf{S}) \geq 0$, as the support displacement is incrementally lowered from $u=0$ at $t=t_0=0$. Such regions are, as expected, localized around the notch tip as well as near the clamped edges of the beam. At each increment, we determine the energy (\ref{Fracture-Energy-Min}) for configurations containing straight through-thickness test cracks, $\Gamma=\Gamma(A,\alpha)$, of crack length $A$, and hence surface area $\mathcal{H}^2(\Gamma) = A D$, emanating from the notch tip and oriented at an angle $\alpha$ relative to the direction of gravity; see Fig.~\ref{Fig4}(a). The restriction to this crack topology is motivated by its consistency with experimental observations. Figure~\ref{Fig4}(b) presents numerical results obtained for the energy $\mathcal{E}$ as a function of $A$ and $\alpha$. The results pertain to $u = 0.52\,\text{mm}$, which is the critical value at which a configuration containing a crack first minimizes (\ref{Fracture-Energy-Min}). The minimizing crack has a length of $A = 7.2\,\text{mm}$ and an orientation of $\alpha = -5^\circ$. This implies that such a crack nucleates at $u = 0.52\,\text{mm}$. The corresponding support force at crack nucleation is $P=206$ N.

\begin{figure}[t!]
\centering
\includegraphics[width=0.95\linewidth]{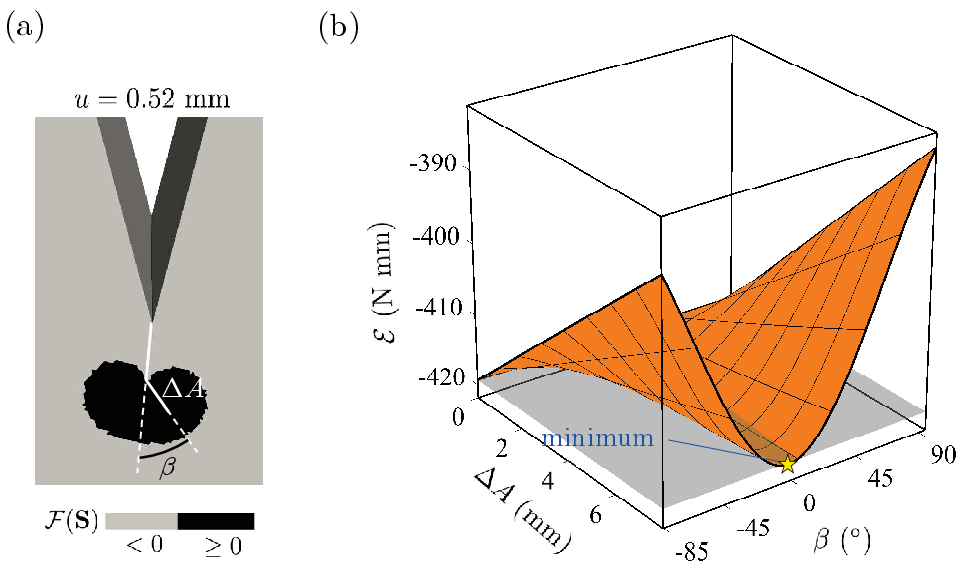}
\caption{{\small Elastic simulations of the test, at the support displacement $u=0.52$ mm, on mortar containing the initial minimizing crack, $(A=7.2\,{\rm mm},\alpha=-5^\circ)$, with test add-cracks of lengths $\Delta A$ oriented at angles $\beta$ with respect to the initial crack orientation. (a) Contour plot of a magnified region of the specimen where the stress field exceeds ($\mathcal{F}(\bfS)\geq0$) the strength surface of the material, depicting the test add-cracks considered within. (b) The energy $\mathcal{E}$ of the specimen plotted as a function of the test add-crack length $\Delta A$ and angle $\beta$.}}\label{Fig5}
\end{figure}

We proceed by analyzing the minimizing solution with crack length $A=7.2\,\text{mm}$ oriented at $\alpha=-5^\circ$. Holding the support displacement at $u = 0.52\,\text{mm}$, we numerically determine the updated region where the stress field violates the strength surface, $\mathcal{F}(\mathbf{S}) \geq 0$. As shown in Fig.~\ref{Fig5}(a), this region of strength violation localizes around the crack front. Within this updated region, we consider straight through-thickness test add-cracks, $\Gamma=\Gamma(A+\Delta A,\alpha,\beta)$, of add-crack length $\Delta A$ oriented at an angle $\beta$ relative to the original crack orientation, resulting in a total crack surface area of $\mathcal{H}^2(\Gamma) = (A + \Delta A) D$. Our energy calculations indicate that $\mathcal{E}$ is minimized for $\Delta A = 7.9\,\text{mm}$ and $\beta = 0^\circ$; see Fig.~\ref{Fig5}(b). Consequently, rather than arresting upon nucleation at $u = 0.52\,\text{mm}$, the crack undergoes finite-size propagation along its original path. Further analogous calculations show that this downward propagation continues until the crack reaches a length of $16.9\,\text{mm}$, at which point it arrests. Lowering the support beyond $u = 0.52\,\text{mm}$ results in additional stable growth of the crack along the same orientation that eventually severs the beam. These predictions agree with the experimental observations.

%
%

\section{Final comments}\label{Sec: Final Comments}

In this Letter, we have introduced a simple benchtop test to directly study fracture driven by gravitational body forces, bypassing the need for specialized testing infrastructure such as centrifuge facilities. Having demonstrated the utility of this approach on a 3D-printable mortar with reproducible results, we anticipate this setup will prove valuable for investigating gravity-driven fracture in other materials relevant to massive structures, such as glacier ice.

The experimental results on the 3D-printable mortar have also served to provide a first validation of the fracture theory recently introduced in \cite{LPKFG2026} for body-force-driven fracture. Specifically, they demonstrate that cracks nucleate and propagate in regions where the strength surface of the material is exceeded, with their evolution dictated by minimizing the sum of the potential and surface energies.

\begin{figure}[t!]
\centering
\includegraphics[width=0.9\linewidth]{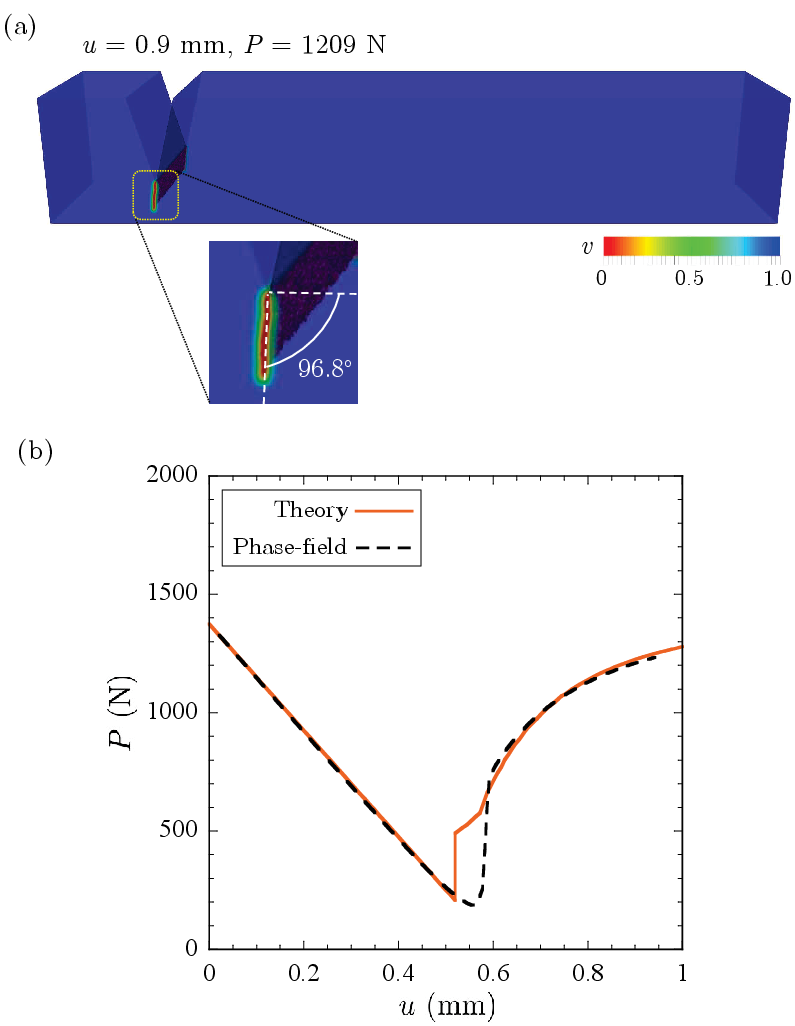}
\caption{{\small (a) Contour plot of the phase field $v$ over the undeformed configuration of the 3D-printable mortar specimen after fracture nucleation and some propagation at $u = 0.9$ mm. (b) Corresponding prediction from the phase-field model (dashed line) compared to the sharp theoretical prediction (solid line) for the support force-displacement ($P$ vs.~$u$) response.}}\label{Fig6}
\end{figure}

The theory introduced in \cite{LPKFG2026} is a sharp theory of fracture that was largely deduced from the regularized phase-field model initiated in \cite{KFLP18} and further elaborated in \cite{KBFLP20,KKLP24,LDLP24}. Although numerical comparisons suggest convergence, a rigorous proof demonstrating that the regularized model converges to the sharp theory in the limit as the regularization length $\varepsilon\searrow 0$ remains an open problem. Howbeit, as shown in Fig.~\ref{Fig6}, which presents the corresponding phase-field predictions\footnote{The phase-field results in Fig.~\ref{Fig6} correspond to a regularization length of $\varepsilon=10$ mm and a mass density $\rho_0$ degraded by $v^2$.} for the test on the 3D-printable mortar, numerical agreement with the sharp theory is maintained in the present case.

%
%

\section*{Acknowledgements}

This work used Expanse at SDSC through allocation MCH260079 from the Advanced Cyberinfrastructure Coordination Ecosystem: Services and Support (ACCESS) program, which is supported by U.S. National Science Foundation grants 2138259, 2138286, 2138307, 2137603, and 2138296. The authors are grateful to Jordan Ouellet and the Newmark Structural Engineering Laboratory for their invaluable assistance in preparing the specimens and carrying out the tests.

\end{document}